\documentclass[12pt]{iopart}

\usepackage{iopams}

\usepackage{graphicx}
\begin{document}
\title
{Ray picture and ray-wave correspondence in triangular microlasers}

\author
{Pia Stockschl\"ader, Jakob Kreismann and Martina Hentschel}

\address
{Institute for Physics, Technische Universit\"at Ilmenau - Weimarer Stra\ss{}e 25, 98693 Ilmenau, Germany}
\ead
{pia.stockschlaeder@tu-ilmenau.de}

\begin{abstract}
 We apply ray-optical methods to dielectric optical microcavities in the shape of triangles made of low refractive index material. 
 We find ray trajectories that maximize the intensity inside the cavity to determine the far-field emission characteristics and to complement the concept of the unstable manifold applicable to chaotic microlasers.
 As these maximum intensity trajectories need not to be periodic, we suggest that they provide a more general explanation for emission patterns of microlasers than short periodic orbits.
 Further, the geometrical optics description is extended by the inclusion of intensity amplification along the optical path to achieve a better description of 
 active, lasing cavities.
 Far-field emission patterns of equilateral triangle cavities obtained in this way agree well with our full electromagnetic wave simulations and with previously reported experimental results.
\end{abstract}

%
%
%
%
%

\section{Introduction}

Dielectric optical microcavities and microlasers have received a lot of interest both as mesoscopic model systems and as devices in micro-optics applications \cite{Vahala,Microcavities_review}. 
They were found, \textit{e.g.}, to show directional far-field emission characteristics \cite{limacon,review_directional}.
One interesting class of those optical billiards are polygonal cavities with triangular resonators as a simple representative.  
In the past, polygonally shaped optical microcavities (modelled often with rounded corners) have been studied both experimentally and theoretically \cite{rounded_isosceles,triangle_MH,Wiersig_BEM,superscars}. 
True polygonal billiards with sharp corners were modelled in detail especially for the closed (hard-wall) case (see \textit{e.g.} \cite{geometry_and_billiards,billiards-in-polygons_1986,billiards-in-polygons_1996,Boshernitzan_polygons} and references therein).

Here, we examine triangular optical billiards with sharp corners as open optical systems. 
It has been shown that the properties of trajectories in generic triangular billiards display a rich behavior depending crucially on the realized geometry \cite{Veech_triangles,equilateral-triangles_periodic,right-triangles_periodic,right-triangles_unstable,nearly_isosceles_triangles}. 

In a recent experiment \cite{triangle_experiments}, various triangular microlasers displaying different symmetries are analyzed. 
The far-field emission patterns of some triangles appear to originate from modes localized on short periodic orbits, whereas the emission of others cannot be explained in this picture. 
We shall see below that indeed another class of orbits, that we will call maximum intensity trajectories, determines and explains the observed far-field emission. 
The experiments are performed with cavities made out of a thin layer of a dye-doped polymer which can be treated as two-dimensional.
The material has a relatively low refractive index of $n < 2$ that corresponds to a rather poor confinement of light by total internal reflection in comparison to typical semiconductor lasers with refractive indices around $n\approx3$. 
Nonetheless, these organic lasers are interesting for applications \cite{organic_dye_lasers,solid-state_organic_lasers}, can be easily processed and optically pumped. 

To gain a better understanding of triangular microlasers with low refractive index, we perform ray-tracing simulations as justified for systems with large size parameters $nkL\gg1$. 
Here $L$ is the characteristic length of the structure, $k=2\pi/\lambda$ the vacuum wavenumber, and $n$ the relative refractive index. 
We include amplification of light relevant in the present case of poor confinement in order to extend the geometrical-optics description to active cavities. 
We also use the wave description for comparison and to analyze the properties of triangular microcavities.

The paper is organized as follows. 
We first discuss the selection of trajectories that contribute to the far-field in the case of the equilateral triangle. 
Further, we examine the influence of light amplification in this system.
Then, we compare the ray optics results to the results of full electromagnetic wave simulations and experimental results.
The simplicity and high symmetry of the equilateral triangle allow us to study the influence of amplification due to an active material in detail and without the obscuring effects of a more complex geometry.

\section{Maximum intensity trajectories}

\begin{figure}
 {\centering
 \includegraphics[width=.9\textwidth]{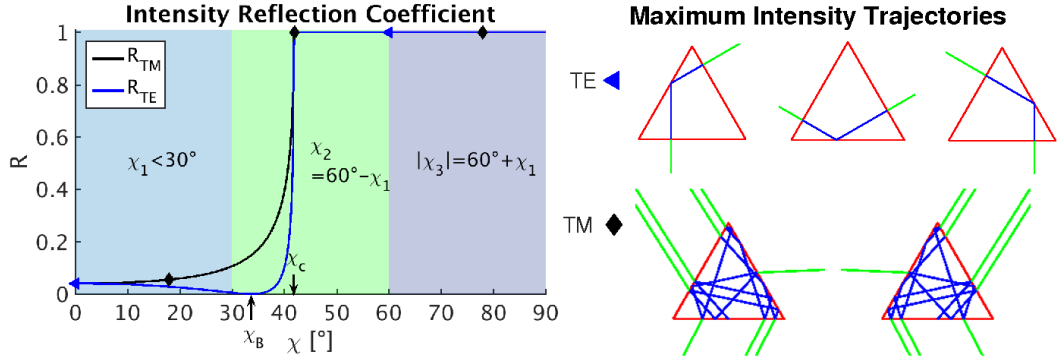}
 
 }
 \caption{\label{fig:orbit_selection}
 \textit{Left:} Intensity reflection coefficient $R(\chi)$ for relative refractive index $n=1.5$ for both polarizations, TE and TM. 
 Critical angle $\chi_{\rm{c}}\approx41.8^\circ$ and Brewster angle (TE) $\chi_{\rm{B}} = \arctan (1/n) \approx33.7^\circ$ are indicated.
 The intervals of possible incident angles in the equilateral triangle are marked by shading. 
 Triangles and diamonds denote the incident angles of the maximum intensity trajectories for TE and TM polarization, respectively.
 \textit{Right:} Examples of the families of maximum intensity trajectories for both polarizations. 
 Quasi-Fabry-Perot orbits with $\chi_1=0^\circ$, $\chi_2=60^\circ$, $\chi_3=-60^\circ$ for TE polarization.
 Non-periodic trajectories with $\chi_1=18^\circ$, $\chi_2=42^\circ$, $\chi_3=-78^\circ$ for TM polarization.
 } 
\end{figure}

The far-field emission of the triangular microlasers studied in \cite{triangle_experiments} could, in many cases, be explained by short and simple periodic orbits. 
One example are the (generalized) Fabry-Perot orbits (\textit{cf.} figure \ref{fig:orbit_selection}) where light hits the resonator boundary vertically on two sides, with a total reflection on the third side in between. 
In order to generalize this picture, and to make a connection to chaotic microlasers where the unstable manifold is known to determine the far-field characteristics \cite{unstable_manifold1,unstable_manifold2,limacon,low-index}, we discuss \textit{all} possible ray trajectories in the equilateral triangle. 
This will enable us to derive a criterion which trajectories contribute to the far-field response. 

A trajectory in a billiard is fully described by its angle of incidence and its position on the boundary at each reflection point.
These two coordinates define the Poincar\'{e} surface of section, a projection of the four-dimensional phase space spanned by the two-dimensional billiard dynamics onto the plane. 
Due to symmetry, each trajectory in an equilateral triangle is characterized by exactly three angles of incidence $\chi_1, \chi_2, \chi_3$. 
Each trajectory is then given by a certain sequence of the $\chi_i$ that depends sensitively on the initial condition (the starting direction chosen at a certain point on one side of the triangle). 
Clearly, a generic trajectory will close only after infinite time and, therefore, is not periodic.  
Let $0^\circ \leq \chi_1 < 30^\circ$ be the initial angle of incidence, then $\chi_2=60^\circ-\chi_1$ and $\chi_3=-(60^\circ+\chi_1)$ \cite{equilateral-triangles_periodic}. 
The sign of the angle $\chi$ specifies the directions of the incoming and outgoing rays at the corresponding reflection point, where the opposite sign changes the direction of the trajectory. 
The angle $\chi_3$ with the largest absolute value always has the opposite sign than the two smaller angles, $\chi_1$ and $\chi_2$. 
Trajectories with reversed signs in all angles are equivalent except for their sense of rotation, we can thus restrict our considerations to the case $\chi_1 \geq 0^\circ$. 
Note that the larger $\chi_1$ is, the less frequent occurs $\chi_3$ along the trajectory sequence. 
The two limiting cases are the ``quasi-Fabry-Perot orbit'' ($\chi_1=0^\circ$ and $\chi_{2/3}=\pm60^\circ$, sequence $\chi_1, \chi_2, \chi_1, \chi_3, \chi_1, \ldots$) and the inscribed triangle (and the corresponding family of period-doubled orbits) with $\chi_1=\chi_2=30^\circ$ where $\chi_3$ does not occur any more. 
 
So far, we have not considered the intrinsic openness of the dielectric cavities.
Hence, we discuss now which of the possible trajectories can be made responsible for the emission characteristics of the dielectric triangular cavity.
We find that trajectories, which maximize the reflected intensity inside the cavity, dominate the far-field emission. 
Although these trajectories are, in general, neither periodic nor simple, they determine the far-field emission for the following reason (maximum intensity trajectory selection rule):  
For the equilateral triangle cavities with relatively low refractive index considered here, at least one of the angles of incidence lies below the critical angle since for $n<2$ the critical angle of total internal reflection is $\chi_{\rm{c}}>30^\circ$. 
Therefore refractive losses are important, and after some initial transition time, trajectories which retain the most reflected intensity will dominate the far-field emission. 
In other words, trajectories are favored that minimize the loss rate along their paths.
This optimization problem has to be solved. 

Now, we apply this method to equilateral triangles with relative refractive index $n=1.5$ yielding a critical angle $\chi_{\rm{c}}\approx41.8^\circ$ and a Brewster angle  (vanishing reflectivity for TE polarization) $\chi_{\rm{B}} = \arctan (1/n) \approx33.7^\circ$.
All following results will be restricted to this refractive index.
The resulting intensity reflection coefficients $R(\chi)$ are depicted in figure \ref{fig:orbit_selection} for both TE and TM polarization. 
For TE polarization, the optimization procedure leads to trajectories with $\chi_1=0^\circ$, $\chi_2=60^\circ$, $\chi_3=-60^\circ$, the so-called ``quasi-Fabry-Perot orbits''.
The maximum intensity trajectories for TM polarization are found to be the family of trajectories with $\chi_1=18^\circ$, $\chi_2=42^\circ$, $\chi_3=-78^\circ$ which are not periodic, as can be seen in figure \ref{fig:orbit_selection}.
For both polarizations, the respective family of trajectories minimizes the losses along their paths independent of the initial position on the boundary under the constraint that the trajectory does not directly hit one of the corners of the triangle.


The large difference between the TM and TE maximum intensity trajectories and, consequently, between the expected far-field emissions 
might be less pronounced in other cavities. 
Each geometry has a particular set of possible trajectories and, hence, specific maximum intensity trajectories depending on the refractive index.
If the maximum intensity trajectories happen to be the same for TE and TM polarizations, the difference between the far-field emission for the two polarizitions is expected to be small.

\section{Amplification in the ray-description}

The usual ray optics simulations follow the rules of classical geometrical optics using the laws of reflection, $\chi_{\rm{ref}}=\chi_{\rm{in}}$, and Snell's law, $\sin(\chi_{\rm{trans}})=n\sin(\chi_{\rm{in}})$, as well as the Fresnel coefficients where $\chi_{\rm{in}}$, $\chi_{\rm{ref}}$ and $\chi_{\rm{trans}}$ are the incident, reflected and transmitted angles, respectively. 
Here, we include amplification along the light path in order to extend the ray model to active, lasing microcavities. 

The reflected and transmitted intensities, $I^{\rm{ref}}$ and $I^{\rm{trans}}$, are obtained from the incident intensity $I^{\rm{in}}$ using the Fresnel equations \cite{Jackson}. 
At the $m$th reflection point of the ray they are given by
\begin{equation}
 I_m^{\rm{ref}} = R(\chi_m) I_m^{\rm{in}} \qquad {\rm{and}} \qquad I_m^{\rm{trans}} = T(\chi_m) I_m^{\rm{in}}
 \label{eq:intensities}
\end{equation}
with the corresponding angle of incidence, $\chi_m$, and the Fresnel reflection and transmission coefficients, $R(\chi_{\rm{in}})$ and $T(\chi_{\rm{in}}) = 1-R(\chi_{\rm{in}})$ that differ for TE and TM polarization, \textit{cf.} figure \ref{fig:orbit_selection}. 
In the case of a passive cavity, the incident intensity is just the reflected intensity of the last bounce, $I_m^{\rm{in}}=I_{m-1}^{\rm{ref}}$. 
We assume no absorption or scattering losses inside the cavity, the only loss mechanism is transmission by refractive escape through the resonator boundary.

To model gain in an active cavity, we assume uniform pumping and a uniform distribution of the active medium throughout the cavity. 
In previous works, this situation was studied within a semiclassical laser theory \cite{laser-theory_Stone} or using the Schr\"odinger-Bloch model 
\cite{sbmodel,spiral_microlasers,gain_MH}. 
A non-uniform gain distribution in chaotic cavities has been studied in the ray model in Ref.~\cite{chaotic_explosions}. 
Generalizing the concept of Husimi functions \cite{MH_husimi_epl} to active cavities illustrated the role of amplification along the light trajectory, and how transmission and reflection of light depend on the previously accumulated intensity \cite{gain_MH}. 
These findings suggest that amplification can be taken into account in an effective manner.

Here, we model the amplification as
\begin{equation}
 I_m^{\rm{in}}=I_{m-1}^{\rm{ref}}\rm{e}^{\alpha \ell_m}
 \label{eq:amp}
\end{equation} 
where $\alpha>0$ is the gain coefficient of the active material and $\ell_m$ is the optical pathlength between the $(m-1)$th and $m$th bounce \cite{Siegman_lasers}. 
This means that the intensity gain is proportional to the intensity $I_{m-1}^{\rm{ref}}$ that enters the piece of trajectory under consideration.
%
In experiments with cavities made of a polymer doped with a laser dye, the above stated assumptions are usually fulfilled. 
Uniform pumping can be obtained when the cavities are optically pumped with the pump beam covering the whole cavity area. 
An approximately uniform distribution of the dye in the polymer matrix is ensured during the liquid phase processing of the material.  
Finally, lasing modes can be assumed to be fully developed even in the case of pulsed pumping as long as the photon round trip time is much shorter than the pump pulse. 
Typical gain coefficients for thin dye-doped polymer layers are of the order of magnitude of $\alpha \sim 10\,\rm{cm}^{-1} \rm{-} 100\,\rm{cm}^{-1}$ \cite{Gozhyk_PRB,Gozhyk_thesis}.

\begin{figure}
 {\centering
 \includegraphics[width=.7\textwidth]{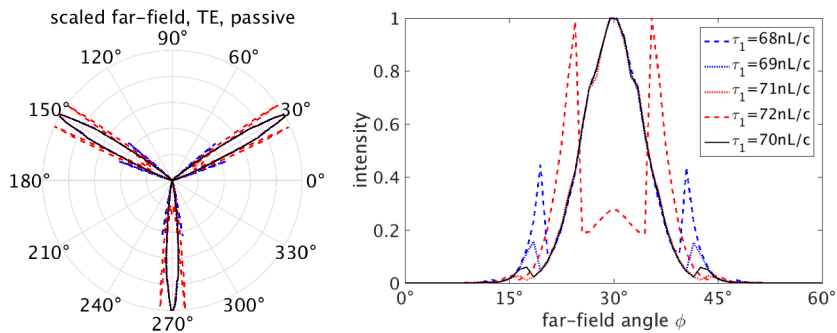}
 
 }
 \caption{\label{fig:farfield_passive}
 Non-universal far-field emission of the equilateral triangle from usual, passive ray optics for TE polarization. 
 The far-field is collected in the time interval $\tau_1 \leq t \leq \tau_2$ with $\tau_1$ varied and $\tau_2 = 86\,nL/c$ for all curves.
 Time is given in units of the optical path length, with the length $L$ of one triangle side and $c/n$ the speed of light in the medium.
 The intensity is scaled such that the maximum equals 1.
 \textit{Left:} Full far-field in polar plot.
 \textit{Right:} Close-up for $0^\circ\leq\phi\leq60^\circ$.
 } 
\end{figure}

Now, we examine the ray-optical calculations in more detail. 
In figure \ref{fig:farfield_passive} the far-field emission from usual passive ray optics is shown 
for a cavity emitting TE polarized light.
To obtain these results, we started $600\,000$ rays ($100\,000$ rays on each side in both directions) with initially unity intensity and random initial conditions uniformly distributed in the angle and the position on the boundary. 
Each trajectory is followed for $100$ reflections.
During this time the total intensity inside the cavity has dropped to less than $10^{-100}$.
To calculate the far-field, the emitted intensities are collected in a time interval $\tau_1 \leq t \leq \tau_2$ where the starting time $\tau_1$ is varied and the end time $\tau_2$ is fixed at the value corresponding to the total length of the shortest trajectory. 

We see that the far-field emission is, indeed, dominated by the predicted maximum intensity trajectories, \textit{i.e.}, for TE polarization the ``quasi-Fabry-Perot orbits'' leading to emission perpendicular to the triangle sides.
We find, however, a strong sensitivity on the time interval chosen to calculate the far-field emission.
%
%
Depending on the chosen starting time $\tau_1$, other directions than those from the predicted maximum intensity trajectories can have a considerable contribution.
After a very long time, we expect the differences to vanish as the family of maximum intensity trajectories will eventually outperform all other trajectories \cite{chaotic_explosions}.
For practical reasons, however, the calculations cannot be done for infinitely long times.
Especially the rapidly decreasing intensities limits the maximum number of reflections for which reasonable and numerically reliable results can be obtained. 
Hence, we cannot deduce a reliable prediction from the passive ray calculations.

\begin{figure}
 {\centering
 \includegraphics[width=.65\textwidth]{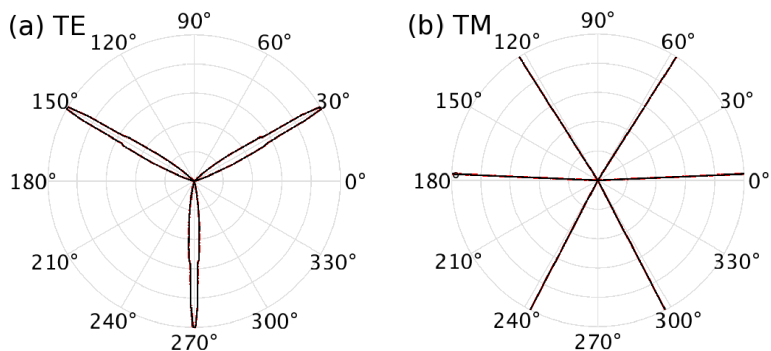}
 
 }
 \caption{\label{fig:farfield_amp}
 Far-field emission of the equilateral triangle from ray optics including amplification according to (\ref{eq:amp}) for (a) TE and (b) TM polarization. 
 The far-field is collected in the same time intervals as in figure \ref{fig:farfield_passive}.
 Obviously, all curves coincide.
 } 
\end{figure}

We find that this problem can be solved if amplification in the active material is included in the ray simulation.
The far-field pattern calculated from the ray model including amplification according to (\ref{eq:amp}) is shown in figure \ref{fig:farfield_amp}.
The gain coefficient is chosen to be $\alpha=3L^{-1}$, where $L$ is the length of one side of the triangle, such that the total intensity inside the cavity increases with time.
All other parameters are the same as in the passive calculation.
Using amplification, the long time limit which is not easily reached in the passive calculation can now be established 
and the influence of the time interval used to collect the far-field intensities is diminished.


For both polarizations, the calculated far-field is now independent of the chosen time interval and can be nicely explained by the predicted families of maximum intensity trajectories.
In the case of TM polarization, the angle of incidence $\chi_{\rm{in}} = 18^\circ$ leads to the angle of transmission $\chi_{\rm{ref}} = \arcsin(n\sin(\chi_{\rm{in}})) \approx 27.6^\circ$. 
Taking into account the threefold symmetry of the cavity and the two possible travelling directions along the trajectory, gives the six observed far-field angles.



\section{Comparison to wave simulations and experimental results}

Next, we compare the ray optics results to results from full electromagnetic wave simulations. 
These simulations are performed with the finite-difference time-domain (FDTD) method \cite{fdtd,fdtd_Taflove}, using a freely available software package \cite{meep}. 
In a first step, the resonant modes of the cavity are calculated.
Then, the far-field emission is determined for the longest-lived modes where the life time is given in terms of the quality factor $Q$.

\begin{figure}
 {\centering
 \includegraphics[width=\textwidth]{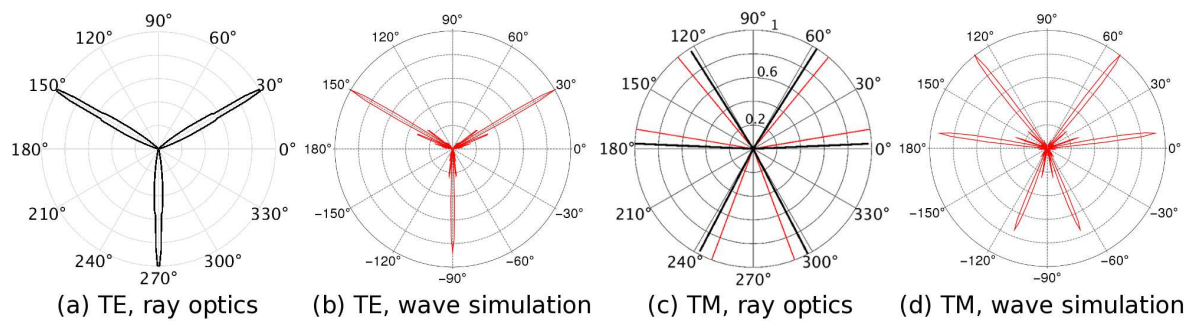}
 
 }
 \caption{\label{fig:far-fields}
 Far-field from ray optics including amplification and from electromagnetic wave simulations for both polarizations.
 Wave results are calculated from the modes with the longest life time found in the equilateral triangle cavity.
 The dimensionless wave numbers and quality factors of the modes are $\rm{Re}(kL)\approx81$, $Q=85$ for TE and $\rm{Re}(kL)\approx96$, $Q=230$ for TM where $k$ is the wavenumber and $L$ is the side length of the triangle.
 The additional red curve in the ray optics result for TM polarization is the result of amended ray optics which accounts 
 for finite wavelength effects.
 } 
\end{figure}

The far-field emission patterns obtained from both approaches, ray optics and wave simulations, are shown in figure \ref{fig:far-fields}.
For TE polarization (first and second panel), we find good agreement of the wave simulations with the ray optics results discussed above.
In both cases, one observes narrow emission peaks perpendicular to the triangle sides as predicted from the maximum intensity trajectories.
The same emission pattern has also been seen in the experiment reported in \cite{triangle_experiments} where only samples with TE polarized lasing emission have been studied.

For TM polarization, however, the agreement between the two approaches is not perfect.
Therefore, we have included semiclassical correction terms in the ray picture in order to account for wave effects \cite{TureciStone02,Aiello_overview,Shinohara2011,SchomerusHentschel_phase-space,PS_EPL2014,Unterhinninghofen_PRE2008}.
The need for wave corrections is obvious as ray optics is strictly only valid in the limit $kL\rightarrow\infty$, 
whereas the wave simulations are performed in the regime $kL\approx100$.
Here, the effect known as Fresnel filtering or angular Goos-H\"anchen effect is og importance and gives corrections to the reflected and transmitted angles, $\chi_{\rm{ref}} = \chi_{\rm{in}} + \Delta\chi_{\rm{ref}}(\chi_{\rm{in}})$ and $\chi_{\rm{trans}} = \arcsin(n\sin(\chi_{\rm{in}})) + \Delta\chi_{\rm{trans}}(\chi_{\rm{in}})$ \cite{TureciStone02,GoetteShinoharaHentschel2013,Microcavities_review,Aiello_overview}.
The black curve in the third panel of figure \ref{fig:far-fields} shows the far-field emission for TM polarization calculated from the ray-optical approach as described before, the red curve is the semiclassically corrected far-field which agrees much better with the wave-optical result. 



The semiclassical correction terms have the largest contribution for incident angles near the critical angle \cite{Aiello_overview,PS_PIERS2015}, therefore, the maximum intensity trajectory for TM polarization with one angle close to the critical angle is strongly affected by the corrections.
Especially, the correction to the transmitted angle is negative \cite{TureciStone02,GoetteShinoharaHentschel2013} which explains the observed change in the far-field for TM polarization if the wave corrections are included in the ray optics (compare black and red curve in figure \ref{fig:far-fields}(c)).
The family of maximum intensity trajectories for TE polarization, however, has incident angles far away from the critical angle, therefore, it is not affected by the corrections and the expected far-field is not changed.
The deviations of the wave simulations from ray optics are assumed to vanish when the limit $kL\rightarrow\infty$ is approached, \textit{e.g.}, when the system size gets larger while keeping the wavelength fixed.

\begin{figure}
 {\centering
 \includegraphics[width=.9\textwidth]{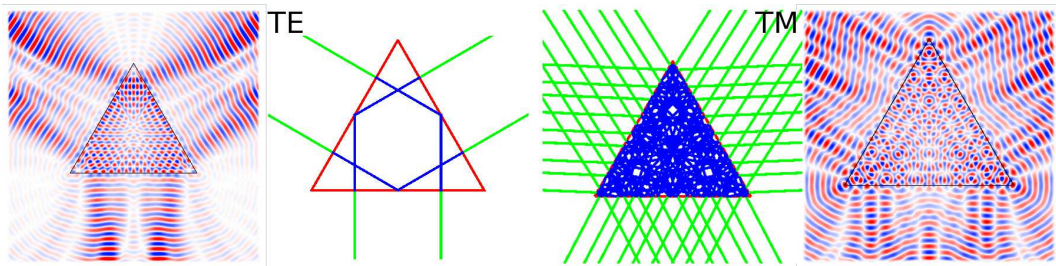}
 
 }
 \caption{\label{fig:modes}
 Mode patterns obtained from full electromagnetic wave simulations in the equilateral triangle cavity in comparison with the predicted maximum intensity trajectories.
 The wave patterns, the magnetic field component $H_z$ for TE polarization and the electric field component $E_z$ for TM polarization, show the modes used to calculate the far-fields in figure \ref{fig:far-fields}.
 } 
\end{figure}

In figure \ref{fig:modes}, the mode patterns of the longest-lived modes of the wave simulations are shown together with examples of the predicted maximum intensity trajectories. 
The qualitative agreement between the wave and ray patterns further illustrates the correspondence of the two approaches and the validity of ray-wave correspondence in triangular microcavities.
The waves which emerge from the sides of the triangle and account for the far-field are clearly visible.
The spherical waves that emerge from the three corners, observed in the wave pattern for TM polarization, indicate diffraction in the near-field (in agreement with the Huygens-Fresnel principle).
However, these diffracted contributions fall-off as (distance)$^{-2}$, thus, they are only visible in the near-field. 
Indeed, they do not leave a trace in the far-field (compare figure \ref{fig:far-fields}(d)). 

\section{Discussion and conclusion}

We have presented a ray optics description of triangular low-refractive-index microlasers confirming ray-wave correspondence.
Including amplification by the active medium into the ray model leads to excellent agreement with full electromagnetic wave simulations and with the experimental results reported in \cite{triangle_experiments}. 
Hence, we conclude that amplification is important to obtain reliable results 
in ray optics simulations of non-chaotic optical cavities, especially in the case of low-index materials and highly lossy geometries.

Often, ray optics has been found a useful model to determine the far-field of microcavities and microlasers, see \textit{e.g.} \cite{Microcavities_review} and references therein.  
In many cases, simple periodic orbits of the classical ray dynamics dominate the spectral properties and the emission, \textit{e.g.} \cite{periodic-orbits_Lebental}.
We find, however, that this assumption is not always sufficient to explain all findings. 
Here, we suggest that maximum intensity trajectories determine the far-field emission characteristics. 
Whereas they may coincide with short periodic orbits as in the case of TE polarization in the equilateral triangle, we demonstrated that this is not generally the case:
We find non-periodic trajectories to dominate the far-field emission of the equilateral triangle cavity for TM polarization.

For classically chaotic systems, the unstable manifold of the chaotic saddle, the way light rays cross the critical line and gets (partially) refracted out of the cavity, was found to determine the far-field characteristics. \cite{unstable_manifold1,unstable_manifold2,limacon,low-index,deformed-cylinders_Schwefel}. 
However, this concept is not applicable to polygonal cavities and other systems having integrable or pseudointegrable classical dynamics. 
The concept of maximum intensity trajectories fills this gap and can provide a more general point of view: 
Trajectories that retain, in the limit of long times, more intensity than others, while still contributing to the transmission, will dominate the far-field. 
This is a similar line of argument as the unstable manifold considerations:
The unstable manifold of a chaotic saddle is constituted by trajectories that undergo many total internal reflections, thus, keeping all their intensity before they are eventually transmitted and contribute to the far-field pattern.
While we have applied the method of maximum intensity trajectories introduced here to the equilateral triangle which has integrable classical dynamics, we assume that this selection rule 
is applicable to 
a large class of cavity and microlaser geometries.

\ack 
 This work is partially funded by the German Research Foundation (DFG) via the Emmy Noether Program.
 The authors thank Joseph Zyss, Stefan Bittner and Melanie Lebental for useful discussions.

\section*{References}

\bibliography{Triangular_Microlasers}

\providecommand{\newblock}{}
\begin{thebibliography}{10}
\expandafter\ifx\csname url\endcsname\relax
  \def\url#1{{\tt #1}}\fi
\expandafter\ifx\csname urlprefix\endcsname\relax\def\urlprefix{URL }\fi
\providecommand{\eprint}[2][]{\url{#2}}

\bibitem{Vahala}
Vahala K 2004 {\em Optical {M}icrocavities\/} (Singapore: World Scientific)

\bibitem{Microcavities_review}
Cao H and Wiersig J 2015 {\em Rev. Mod. Phys.\/} {\bf 87}(1) 61--111

\bibitem{limacon}
Wiersig J and Hentschel M 2008 {\em Phys. Rev. Lett.\/} {\bf 100}(3) 033901

\bibitem{review_directional}
Wiersig J, Unterhinninghofen J, Song Q, Cao H, Hentschel M and Shinohara S 2011
  Review on unidirectional light emission from ultralow-loss modes in deformed
  microdisks {\em Trends in Nano- and Micro-Cavities\/} ed Kwon O, Lee B and An
  K (Bentham Science Publishers) pp 109--152

\bibitem{rounded_isosceles}
Kurdoglyan M~S, Lee S~Y, Rim S and Kim C~M 2004 {\em Opt. Lett.\/} {\bf 29}
  2758--2760

\bibitem{triangle_MH}
Hentschel M, Wang Q~J, Yan C, Capasso F, Edamura T and Kan H 2010 {\em Opt.
  Express\/} {\bf 18} 16437--16442

\bibitem{Wiersig_BEM}
Wiersig J 2003 {\em Journal of Optics A: Pure and Applied Optics\/} {\bf 5} 53

\bibitem{superscars}
Bogomolny E, Dietz B, Friedrich T, Miski-Oglu M, Richter A, Sch\"afer F and
  Schmit C 2006 {\em Phys. Rev. Lett.\/} {\bf 97}(25) 254102

\bibitem{geometry_and_billiards}
Tabachnikov S 2009 {\em Geometry and billiards\/} 2nd ed (Providence, RI:
  American Mathematical Society)

\bibitem{billiards-in-polygons_1986}
Gutkin E 1986 {\em Physica D: Nonlinear Phenomena\/} {\bf 19} 311--333

\bibitem{billiards-in-polygons_1996}
Gutkin E 1996 {\em Journal of Statistical Physics\/} {\bf 83} 7--26

\bibitem{Boshernitzan_polygons}
Boshernitzan M, Galperin G, Kr{\"u}ger T and Troubetzkoy S 1998 {\em
  Transactions of the American Mathematical Society\/} {\bf 350} 3523--3535

\bibitem{Veech_triangles}
Veech W 1989 {\em Inventiones mathematicae\/} {\bf 97} 553--583

\bibitem{equilateral-triangles_periodic}
Baxter A~M and Umble R 2008 {\em The American Mathematical Monthly\/} {\bf 115}
  479--491

\bibitem{right-triangles_periodic}
Cipra B, Hanson R~M and Kolan A 1995 {\em Phys. Rev. E\/} {\bf 52}(2)
  2066--2071

\bibitem{right-triangles_unstable}
Hooper W 2007 {\em Geometriae Dedicata\/} {\bf 125} 39--46

\bibitem{nearly_isosceles_triangles}
Hooper W~P and Schwartz R~E 2009 {\em Journal of Modern Dynamics\/} {\bf 3}
  159--231

\bibitem{triangle_experiments}
Lafargue C, Lebental M, Grigis A, Ulysse C, Gozhyk I, Djellali N, Zyss J and
  Bittner S 2014 {\em Phys. Rev. E\/} {\bf 90}(5) 052922

\bibitem{organic_dye_lasers}
Soffer B~H and McFarland B~B 1967 {\em Applied Physics Letters\/} {\bf 10}
  266--267

\bibitem{solid-state_organic_lasers}
Ch{\'e}nais S and Forget S 2012 {\em Polymer International\/} {\bf 61} 390--406

\bibitem{unstable_manifold1}
Lee S~Y, Ryu J~W, Kwon T~Y, Rim S and Kim C~M 2005 {\em Phys. Rev. A\/} {\bf
  72}(6) 061801

\bibitem{unstable_manifold2}
Shinohara S and Harayama T 2007 {\em Phys. Rev. E\/} {\bf 75}(3) 036216

\bibitem{low-index}
Schermer M, Bittner S, Singh G, Ulysse C, Lebental M and Wiersig J 2015 {\em
  Applied Physics Letters\/} {\bf 106} 101107

\bibitem{Jackson}
Jackson J~D 1999 {\em Classical electrodynamics\/} 3rd ed (New York, NY: Wiley)

\bibitem{laser-theory_Stone}
T\"ureci H~E, Stone A~D and Collier B 2006 {\em Phys. Rev. A\/} {\bf 74}(4)
  043822

\bibitem{sbmodel}
Harayama T, Davis P and Ikeda K~S 2003 {\em Phys. Rev. Lett.\/} {\bf 90}(6)
  063901

\bibitem{spiral_microlasers}
Hentschel M and Kwon T~Y 2009 {\em Opt. Lett.\/} {\bf 34} 163--165

\bibitem{gain_MH}
Kwon T~Y, Lee S~Y, Ryu J~W and Hentschel M 2013 {\em Phys. Rev. A\/} {\bf
  88}(2) 023855

\bibitem{chaotic_explosions}
Altmann E~G, Portela J~S~E and T\'{e}l T 2015 {\em EPL (Europhysics Letters)\/}
  {\bf 109} 30003

\bibitem{MH_husimi_epl}
Hentschel M, Schomerus H and Schubert R 2003 {\em EPL (Europhysics Letters)\/}
  {\bf 62} 636

\bibitem{Siegman_lasers}
Siegman A~E 1986 {\em Lasers\/} (Mill Valley, CA: Univ. Science Books)

\bibitem{Gozhyk_PRB}
Gozhyk I, Boudreau M, Haghighi H~R, Djellali N, Forget S, Ch\'enais S, Ulysse
  C, Brosseau A, Pansu R, Audibert J~F, Gauvin S, Zyss J and Lebental M 2015
  {\em Phys. Rev. B\/} {\bf 92}(21) 214202

\bibitem{Gozhyk_thesis}
Gozhyk I 2012 {\em {Polarization and gain phenomena in dye-doped polymer
  micro-lasers}\/} Phd thesis {{\'E}cole normale sup{\'e}rieure de Cachan}

\bibitem{fdtd}
Inan U~S and Marshall R~A 2011 {\em Numerical Electromagnetics: The FDTD
  Method\/} (Cambridge: Cambridge University Press)

\bibitem{fdtd_Taflove}
Taflove A and Hagness S~C 2005 {\em Advances in Computational Electrodynamics:
  The Finite-difference Time-domain Method\/} 3rd ed (Boston, MA: Artech House)

\bibitem{meep}
Oskooi A~F, Roundy D, Ibanescu M, Bermel P, Joannopoulos J~D and Johnson S~G
  2010 {\em Computer Physics Communications\/} {\bf 181} 687--702

\bibitem{TureciStone02}
Tureci H~E and Stone A~D 2002 {\em Opt. Lett.\/} {\bf 27} 7--9

\bibitem{Aiello_overview}
Bliokh K~Y and Aiello A 2013 {\em J. Opt.\/} {\bf 15} 014001

\bibitem{Shinohara2011}
Shinohara S, Harayama T and Fukushima T 2011 {\em Opt. Lett.\/} {\bf 36}
  1023--1025

\bibitem{SchomerusHentschel_phase-space}
Schomerus H and Hentschel M 2006 {\em Phys. Rev. Lett.\/} {\bf 96}(24) 243903

\bibitem{PS_EPL2014}
Stockschl{\"{a}}der P, Kreismann J and Hentschel M 2014 {\em EPL (Europhysics
  Letters)\/} {\bf 107} 64001

\bibitem{Unterhinninghofen_PRE2008}
Unterhinninghofen J, Wiersig J and Hentschel M 2008 {\em Phys. Rev. E\/} {\bf
  78}(1) 016201

\bibitem{GoetteShinoharaHentschel2013}
G{\"o}tte J~B, Shinohara S and Hentschel M 2013 {\em J. Opt.\/} {\bf 15} 014009

\bibitem{PS_PIERS2015}
Stockschl\"{a}der P, Kreismann J and Hentschel M 2015 Wave-inspired corrections
  for an efficient ray-optical description of micro-optics devices {\em PIERS
  Prague 2015 Proceedings\/} (Cambridge, MA: The Electromagnetics Academy) pp
  1647--1651

\bibitem{periodic-orbits_Lebental}
Lebental M, Djellali N, Arnaud C, Lauret J~S, Zyss J, Dubertrand R, Schmit C
  and Bogomolny E 2007 {\em Phys. Rev. A\/} {\bf 76}(2) 023830

\bibitem{deformed-cylinders_Schwefel}
Schwefel H~G~L, Rex N~B, Tureci H~E, Chang R~K, Stone A~D, Ben-Messaoud T and
  Zyss J 2004 {\em J. Opt. Soc. Am. B\/} {\bf 21} 923--934

\end{thebibliography}

\end{document}